\documentclass[a4paper,12pt,onecolumn,fleqn,twoside]{article}

\usepackage[margin=3.0cm]{geometry}

\usepackage{times}

\usepackage[small,bf]{caption}
\usepackage{graphicx}

\usepackage[square,numbers,sort&compress]{natbib}
\usepackage{epsfig}
\usepackage[runin]{abstract}

\usepackage{latexsym}
\usepackage{amsmath}
\usepackage{amssymb}
\usepackage{bbm}

\allowdisplaybreaks[4] 

   \newcommand{\expvaluePsi}[1]{\ensuremath{\left \langle \Psi \left | #1 \right | \Psi \right \rangle}}
   
   \newcommand{\expvalueGeneral}[2]{\ensuremath{\left \langle #2 \left | #1 \right | #2 \right \rangle}}
   
   \newcommand{\condensate}[1]{\ensuremath{\left \langle #1 \right \rangle}}
   
   \newcommand{\tkappa}[1]{\ensuremath{\tilde{\kappa}}}
   
\begin{document}

\pagestyle{empty}

\title{Hunting Medium Modifications\\of the Chiral Condensate}

\author{R. Thomas$^{a}$, S. Zschocke$^{a,c}$, T. Hilger$^{a,b}$, B. K\"ampfer$^{a,b}$ \\
\small $^a$ Institut f\"ur Kern- und Hadronenphysik, Forschungszentrum Rossendorf, \vspace{-5pt} \\
\small PF 510119, 01314 Dresden, Germany \vspace{-5pt} \\
\small $^b$ Institut f\"ur Theoretische Physik, TU Dresden, 01062 Dresden, Germany \vspace{-5pt} \\
\small $^c$ Theoretical and Computational Physics Section, University of Bergen, \vspace{-5pt} \\
\small 5007 Bergen, Norway \vspace{-5pt}
}

\date{}

\maketitle
\thispagestyle{empty}

\renewcommand{\abstractname}{}
\begin{abstract}
With QCD sum rule evaluations, spectral changes of hadrons inside nuclear matter are considered, which shed light on QCD condensates and thus on the non-perturbative structure of the QCD ground state. For some light quark configurations, $\omega$ meson and nucleon, the relevance of four-quark condensates is compared; the $D$ meson as a representative of heavy-light quark systems is also briefly discussed.
\end{abstract}

\section{Introduction}
Hadrons are confined composite systems built out of colored quarks and 
described by the theory of strong interaction.
These hadrons are excited from the ground state of the theory.
A change in properties of this ground state is expected to reflect in a change of hadronic properties, especially in spectral functions and moments
thereof related to masses of hadrons.
Brown and Rho~\cite{Brown:1991kk} argued that for light vector mesons a scaling law like $m_V \sim \condensate{\bar{q} q}^x$ holds, where $\condensate{\bar{q} q}$ is the chiral condensate and measures the spontaneous breaking of chiral symmetry and $x$ denotes some positive power. Measurements of ''mass modifications'' of hadrons at finite temperature or when situated inside nuclear matter, that
means embedded in a bulk of protons and neutrons, could then probe the QCD vacuum. The properties of this ground state are quantified in a number of condensates, which, like universal material constants, carry important information on symmetry features of the theory. Besides $\condensate{\bar{q} q}$ up to mass dimension 6 the gluon condensate $\condensate{\tfrac{\alpha_s}{\pi} G^2 }$, the mixed quark-gluon condensate $\condensate{\bar{q} g_s \sigma G q}$, the triple gluon condensate $\condensate{g_s^3 G^3}$ and structures of the form $\condensate{\bar{q} \Gamma q \bar{q} \Gamma q}$ contribute. ($\Gamma$ denotes all possible structures formed by Dirac and Gell-Mann matrices.) We emphasize here the specific role of the latter class of condensates, which still lack information, the four-quark condensates.

We discuss how some of these condensates could change under the influence of 
surrounding nuclear matter. Therefore, the method of QCD sum rules is applied, which is able to deal with a large variety of
hadrons, so our findings and similar considerations for other hadrons can reveal more insight into changes of hadronic properties
at finite baryon density.

\section{QCD Sum Rules at Finite Density}
The effect of the non-perturbative regime of QCD can be accounted for by including power corrections to the correlation function
\begin{equation}
\Pi_{\mu \nu} (q) = i \int d^4 x \, e^{iqx} \, \expvaluePsi{T[j_\mu (x) j_\nu (0)]}
\label{eq:correlationFunction}
\end{equation}
of the hadron described by the current $j_\mu$. These corrections for large space-like momenta $q^2$ enter via an operator product expansion (OPE) of~(\ref{eq:correlationFunction}) and introduce Wilson coefficients multiplied by local normal ordered expectation values of quark and gluon fields -- the QCD condensates. On the other side the correlation function is related solely to the properties of the hadronic degrees of freedom, often taking a simple pole ansatz of a low-lying resonance or, more generally, considering moments of spectral distribution functions valid for $q^2 > 0$. Due to the assumed analyticity of the correlator $\Pi_{\mu \nu}$, dispersion relations equate both approaches. This leads to the celebrated QCD sum rules introduced by Shifman, Vainshtein and Zakharov~\cite{Shifman:1978bx}. In order to draw conclusions for the hadron masses, say, an appropriate range in $q^2$ has to be considered; various evaluation procedures exist. We work with Borel transformed dispersion relations where unknown polynomials are cancelled, higher excitations in the hadronic ansatz are suppressed and the convergence of the operator product expansion is improved.

\begin{figure}[htb]
\begin{center}
\vspace*{-0.0cm}
\includegraphics[width=5cm,angle=270]{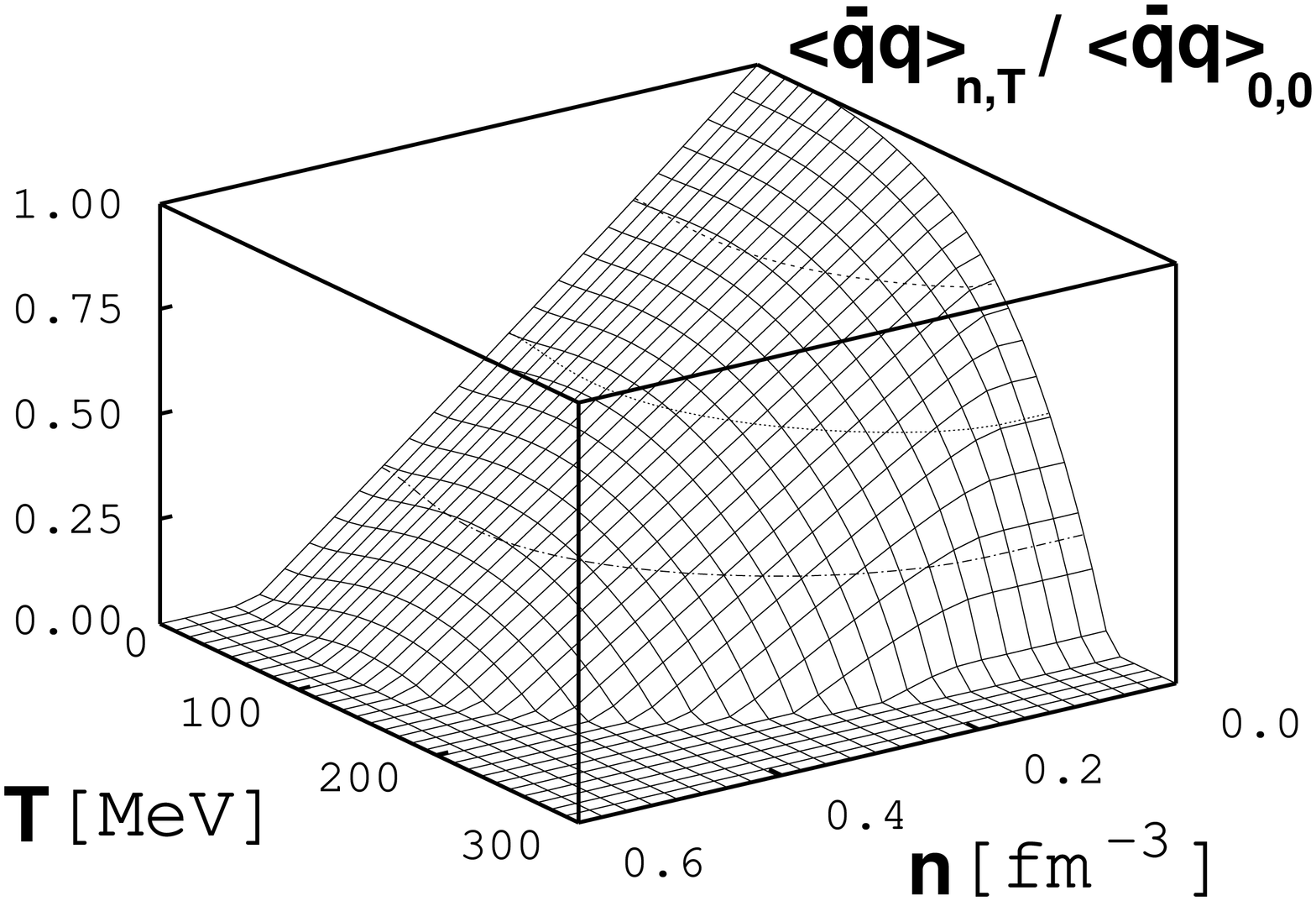}
\includegraphics[width=5cm,angle=270]{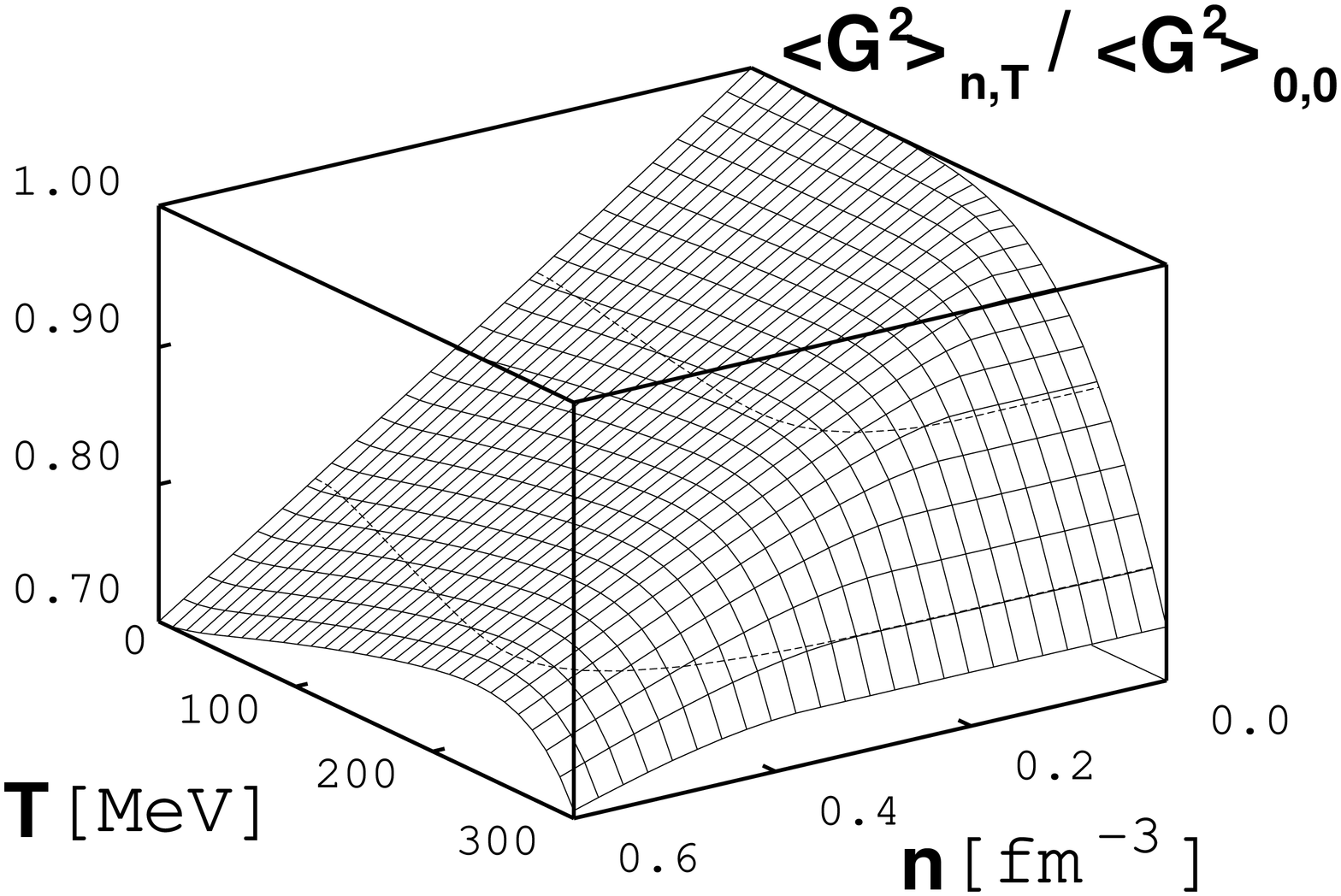}
\vspace*{-0.0cm}
\caption{The chiral condensate (left panel) and gluon condensate (right panel) as functions of temperature $T$ and density $n$ in dilute gas approximation~\cite{Zschocke:2002ic}.}
\label{fig:condensatesDensityTemperature}
\end{center}
\end{figure}

Effects of finite baryon density $n$ or temperature $T$ are described by the change of condensates and the advent of new structures which are absent at zero density or temperature. In dilute gas approximation the medium changes are linear in densities as evidenced by the leading terms
\begin{equation}
\condensate{\mathcal{O}} = \condensate{\mathcal{O}}_0 + \dfrac{n}{2M_N} \expvalueGeneral{\mathcal{O}}{N} + \dfrac{T^2}{8} \expvalueGeneral{\mathcal{O}}{\pi} + \ldots \,.
\end{equation}
For instance, Fig.~\ref{fig:condensatesDensityTemperature} shows that with increasing temperature or density the chiral and gluon condensates decrease. If one relies on Brown-Rho scaling this implies a lowering of masses of light vector mesons~\cite{Hatsuda:1991ez}, a question experimentally looked for by HADES in the dielectron decay channel~\cite{Friese:1999qm}. Even the often quoted Ioffe formula for the nucleon mass $M_N \propto \condensate{\bar{q} q}$ points to similar conclusions. However, instead of simple pocket formulas, one has to perform consistent QCD sum rule evaluations for an accurate inclusion of relevant condensates. Therefore, we consider here examples of different light or heavy quark configurations: $\omega$, $N$, $D$ and $J/\Psi$; and focus on baryon density effects.

\section{What Medium Modifications Tell about Condensates?}
\subsection{Light quark vector mesons: The case of \boldmath{$\omega$} }
In sum rules for light vector mesons the behavior of the $\omega$ in-medium mass is dominated by the density dependence of a combination of four-quark condensates \cite{Zschocke:2002mp,Thomas:2005wc}. The CB-TAPS collaboration observed in the reaction 
$\gamma + {\rm A} \to {\rm A}' + \omega (\to \pi^0 \gamma)$ the occurrence of 
additional low-energy $\omega$ decay strength for a Nb target compared 
to a hydrogen target \cite{Trnka:2005ey}. This can be used to constrain the density behavior of the specific four-quark condensates in the $\omega$ sum rule \cite{Thomas:2005dc}. Writing out the sum rule equation for the first moment 
of the spectral distribution function leads to
\begin{equation}
m_\omega^2 (n,{\cal M}^2,s_\omega) =
\frac{c_0 {\cal M}^2 
\left[ 1 - \left ( 1 + \frac{s_\omega}{{\cal M}^2} \right) e^{-s_\omega / {\cal M}^2} \right] - 
\frac{c_2}{{\cal M}^2} - \frac{c_3}{{\cal M}^4} - \frac{c_4}{2 {\cal M}^6}}
{c_0 \left ( 1 - e^{-s_\omega / {\cal M}^2} \right) + \frac{c_1}{{\cal M}^2} 
+ \frac{c_2}{{\cal M}^4} + 
\frac{c_3}{2 {\cal M}^6} + 
\frac{c_4}{6 {\cal M}^8} - 
\frac{\Pi^\omega (0,n)}{{\cal M}^2}} \,.
\label{eq:massEquationOmega}
\end{equation}
Here, ${\cal M}$ is the Borel mass, $c_{j}$ are the terms of the OPE including Wilson coefficients and condensates
(for details cf.~\cite{Zschocke:2004qa}), and $s_\omega$ is the so-called continuum threshold. The moment
$m_\omega (n,{\cal M}^2,s_\omega)$ is defined as
\begin{equation}
m_\omega^2 (n,{\cal M}^2,s_\omega) \equiv \frac{\int_0^{s_\omega} ds \; {\rm Im} \Pi^\omega (s,n) \; 
e^{-s/{\cal M}^2}}{\int_0^{s_\omega} ds \; {\rm Im} \Pi^\omega (s,n) \; s^{-1} e^{-s/{\cal M}^2}} \, ;
\label{eq:firstMomentDefinition}
\end{equation}
for a pole ansatz for ${\rm Im} \Pi^\omega$ it
corresponds to the pole mass.
Especially in $c_3$ the flavor-mixing four-quark condensates 
$\frac29 \langle \bar u \gamma^\mu \lambda_A u \bar d \gamma_\mu \lambda_A d \rangle
+
\langle \bar u \gamma_5 \gamma^\mu \lambda_A u 
\bar d \gamma_5 \gamma_\mu \lambda_A d \rangle$
and the pure flavor four-quark condensates (for which we employ $u - d$ isospin symmetry;
$\gamma_\mu$ and $\lambda_A$ stand for Dirac and Gell-Mann matrices)
$\frac29 \langle \bar q \gamma^\mu \lambda_A q \bar q \gamma_\mu \lambda_A q \rangle
+
\langle \bar q \gamma_5 \gamma^\mu \lambda_A q 
\bar q \gamma_5 \gamma_\mu \lambda_A q \rangle$ enter.
The density dependence of these condensates is taken as the linearized density behavior of the squared chiral condensate multiplied by parameters $\kappa$.
For the given combination of four-quark condensates an effective parameter $\kappa_N$ describes the strength of their total density dependence.
\begin{figure}[htb]
\begin{center}
\vspace*{+0.3cm}
\includegraphics[width=6.5cm,angle=270]{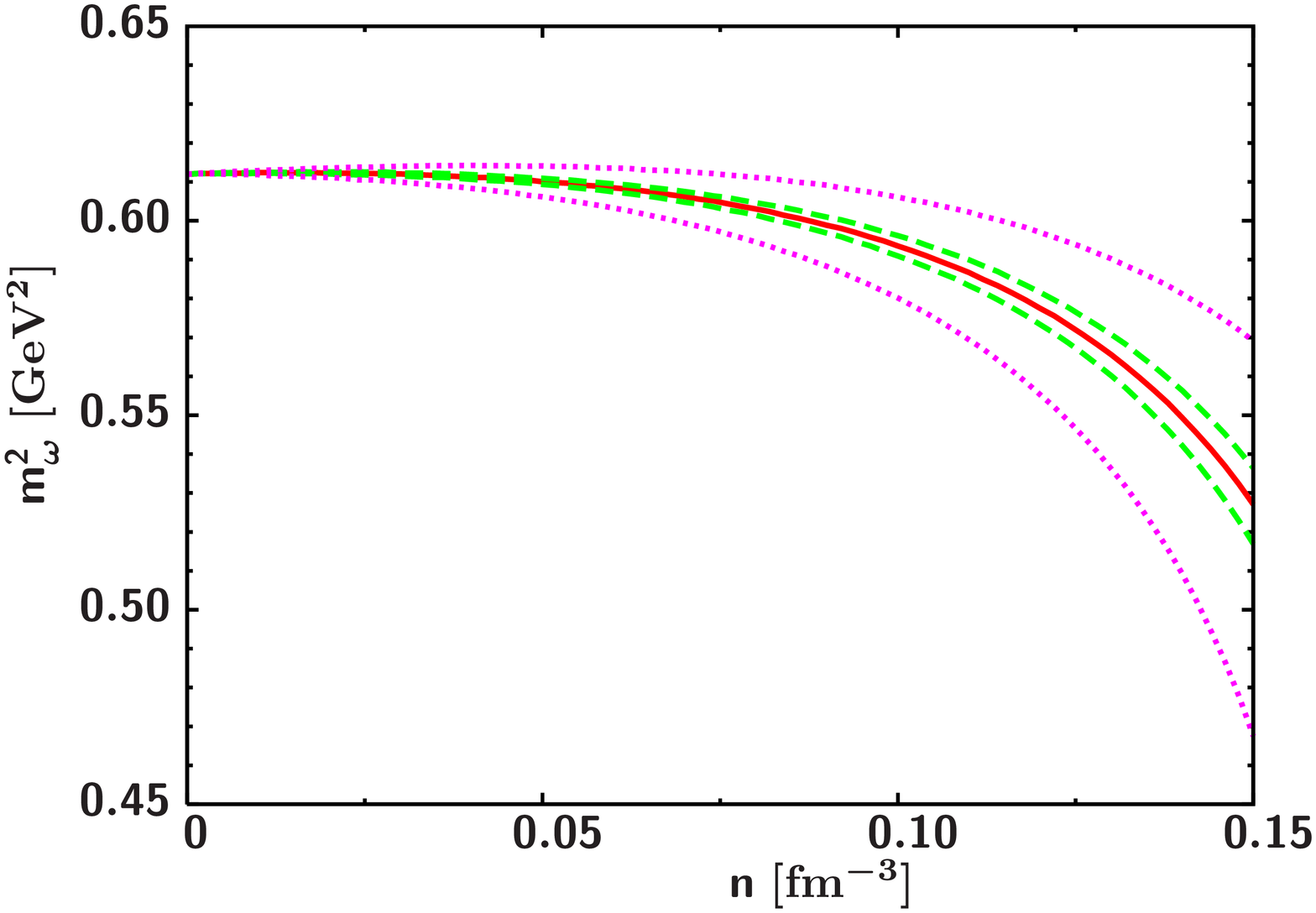}
\vspace*{-0.5cm}
\caption{The mass parameter $m_\omega^2$ from a full QCD sum rule evaluation 
as a function of the baryon density
for $\kappa_N = 4$ and $c_4 = 0$ (solid curve).
The effect of a $c_4^{(1)}$ term
in the expansion $c_4 = c_4^{(0)} + c_4^{(1)} n / n_0$
is exhibited, too
($c_4^{(1)} = \pm 10^{-5} n_0^{-1}$ GeV\,$^8$: dashed curves,
$c_4^{(1)} = \pm 5 \times 10^{-5} n_0^{-1}$ GeV\,$^8$: dotted curves;
the upper (lower) curves are for negative (positive) signs).
For further details cf.~\cite{Thomas:2005dc}.}
\label{fig:omegaDensityPRL}
\end{center}
\end{figure}
The in-medium modification of the light vector meson masses crucially depends on the density dependence of four-quark condensates, which can already be estimated analytically if one expands the given sum rule equation for small densities and employs standard condensate values and parameters from the full sum rule evaluation
\begin{equation}
m_\omega^2 (n) = m_\omega^2 (0) + n (4 -\kappa_N) \frac{0.03}{n_0} \,{\rm GeV}^2 \, .
\end{equation}
The estimate points out that a strong density dependence $(\kappa_N > 4)$ 
is required for an at least not increasing moment $m_\omega^2$.
(Note that QCD sum rules do not specify the shape of the in-medium spectral functions, but constrain moments thereof.)

In Fig.~\ref{fig:omegaDensityPRL} this critical situation is exhibited, further, the inclusion of
the next higher dimensional condensates is shown. It changes the absolute value of the critical $\kappa_N$, but not the qualitative requirement of a strong density dependence of the combined four-quark condensates for consistency with the observation of CB-TAPS. This implies at finite densities a considerable drop of the four-quark condensate combination considered here.
As a consequence the integrated $\rho$ spectral density must be ''down shifted'' even stronger.

\subsection{Light quark baryons: Nucleon}
We now apply QCD sum rules to derive predictions on the in-medium behavior of self-energies of the nucleon as they occur in its propagation function
\begin{equation}
G(q) = \frac{1}{q_\mu \gamma^\mu - M_\mathrm{N} - \Sigma (q)} \, ,
\end{equation}
which can be decomposed into invariant parts using the two characteristic Lorentz vectors of the problem, the nucleon 
momentum $q_\mu$ and the medium velocity $v_\mu$. This decomposition together with 
the pole structure of $G(q)$ leads to the definition of the scalar and vector self-energies $\Sigma_{\mathrm s}$ and 
$\Sigma_\mathrm{v}$. The first accounts for the deviation from the nucleon vacuum mass $M_{\mathrm N}$. In-medium chiral 
perturbation theory for example predicts large cancellation effects between these two quantities \cite{Vretenar:2003bt}, i.e. the 
self-energies should be comparable in magnitude but have opposite signs.

\begin{figure}[htb]
\begin{center}
\vspace*{+0.3cm}
\includegraphics[width=6.5cm,angle=270]{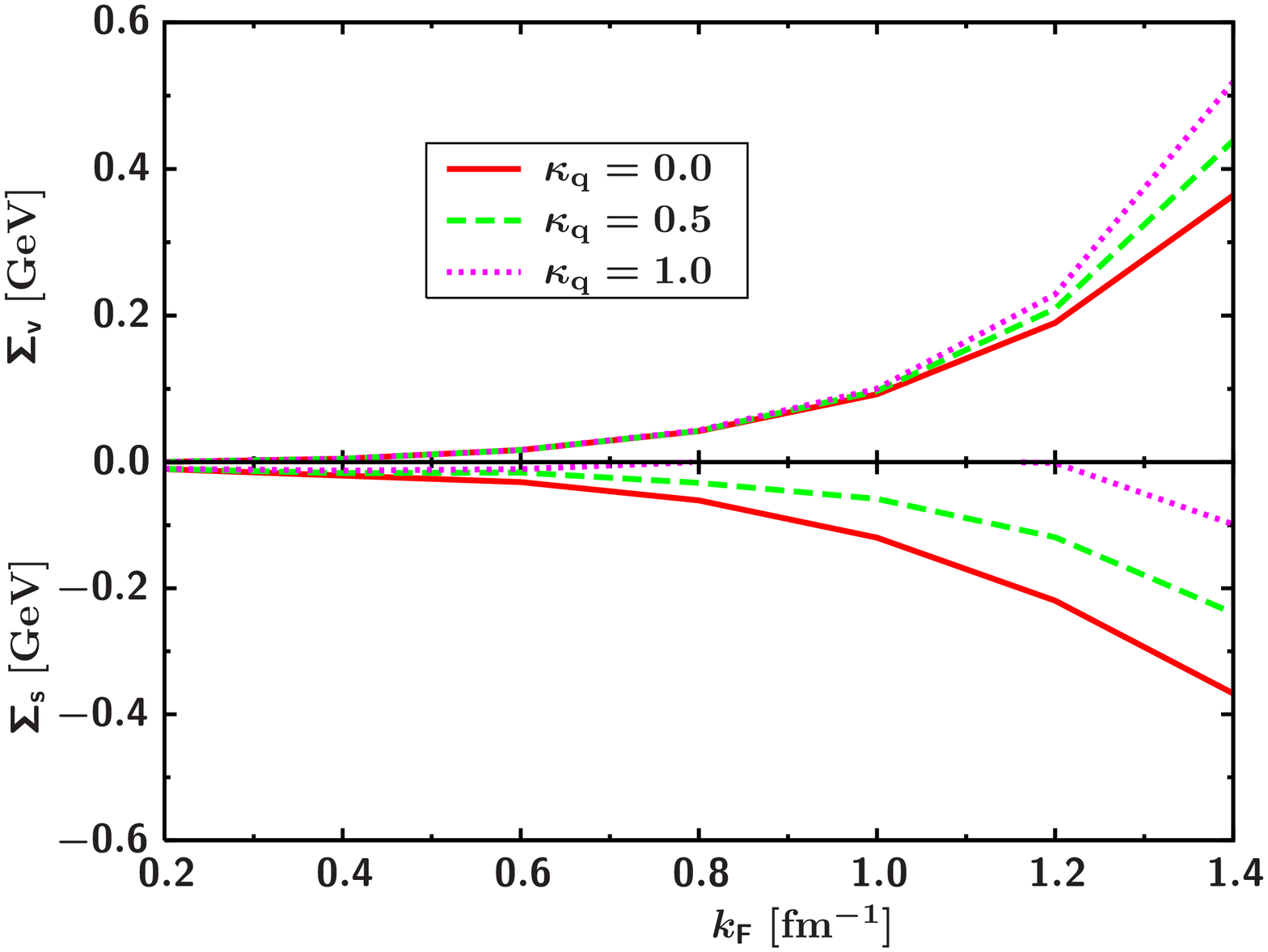}
\vspace*{-0.6cm}
\caption{The scalar and vector self-energies $\Sigma_\mathrm{s}$ (upper part) and $\Sigma_\mathrm{v}$ (lower part) of the nucleon as functions of the 
nucleon Fermi momentum $k_\mathrm{F}$ for different density dependencies of the combined four-quark condensates. The 
preferred symmetric picture is realized if this dependence is weak.}
\label{fig:nucleonDensity}
\end{center}
\end{figure}

QCD sum rule evaluations showed a significant dependence of the self-energies on the density behavior of four-quark 
condensates \cite{Furnstahl:1992pi}. The standard treatment for four-quark condensates is to simplify it to squares of the chiral condensate 
$\langle \bar{q} q \rangle$ (this is the ground state saturation approximation), which is supported by large $N_c$ arguments. 
As for the $\omega$ case above we repeat to test this approximation while we apply our parameterization of four-quark condensates for the full set of structures which appear in the operator product expansion for the nucleon interpolating fields. The sum of all parameterized 
four-quark condensates up to a linear dependence in baryon density $n$
$$
\sum_{\mathrm{\Gamma,C,f1,f2}}
\alpha_{\langle \cdot \rangle} \langle \bar{q}_{\mathrm f1} \Gamma_1 
{\mathrm C_1} q_{\mathrm f1} \bar{q}_{\mathrm f2} \Gamma_2 {\mathrm C_2} 
q_{\mathrm f2} \rangle = A \kappa^\mathrm{vac}_\mathrm{N} + 
B \kappa^\mathrm{med}_\mathrm{N} n \nonumber
$$
yields effective parameters $\kappa^\mathrm{vac}_\mathrm{N}$ and $\kappa^\mathrm{med}_\mathrm{N}$ which describe 
deviations from the standard treatment.
The sum runs over all four-quark condensates with weight factors $\alpha_{\langle \cdot \rangle}$ and extends the 
condensate list for the vacuum case \cite{Thomas:2005wc}, $\Gamma, \mathrm{C}$ and $\mathrm{f}$ denote Dirac, color and flavor structures; 
$A$ and $B$ are constants and estimate the order of magnitude as given by the factorization limit.
A value of $\kappa^\mathrm{med}_\mathrm{N}=0$ corresponds to no density-dependence of the combined four-quark condensates, 
a value of $\kappa^\mathrm{med}_\mathrm{N} =\kappa^\mathrm{vac}_\mathrm{N}$ to a density behavior similar to that of the 
squared chiral condensate ($\kappa^\mathrm{vac}_\mathrm{N}$ is adapted to the correct vacuum nucleon mass). The results 
in Fig.~\ref{fig:nucleonDensity} exhibit the variation of the dominant $\kappa_\mathrm{q}$ and its consequences for the self-energies as function 
of the Fermi momentum of the nucleon. The sum rules for the nucleon involve actually three equations each of them 
containing its own set of four-quark condensates, we consider here the most sensitive part with $\kappa_\mathrm{q}$. A 
symmetric picture of self-energies is achieved if the combined four-quark condensates are hardly density dependent, expressed by 
$\kappa_\mathrm{q} \approx 0$. Calculations of four-quark condensates in a perturbative chiral quark model support this choice 
\cite{Drukarev:2003xd}.

The four-quark condensates in sum rules for baryons like the nucleon differ significantly from those in meson sum rules. 
The three quark systems generically yield the occurrence of linear combinations of these condensates already in zeroth order of $\alpha_s$
in two different color structures.
This interestingly leads to a reduction of the list of independent four-quark structures for baryons, 
contrary it makes a connection to meson four-quark condensates partially impossible, since the latter appear only with 
one type of color contraction. Therefore, a consideration of structures in other baryon sum rules, for example the 
$\Delta$, could provide more information on four-quark condensates and especially how the sum rules of nucleon and other 
baryons are related in this respective.

\subsection{Heavy quark systems: \boldmath{$D$} meson and \boldmath{$J/\Psi$} }
In confined systems including a heavy quark $Q$ (say charm) the inverse mass deals as expansion parameter, further other condensates can become important.
For example, Wilson coefficients which contain the heavy quark mass like a term $m_Q \condensate{\bar{q} q}$ are numerically amplified (contrary to the
$\omega$ sum rule). In the case of charmed mesons one can relate combinations of positive and negative frequency parts of the correlation function to even and odd components with respect to $q_0$ of the operator product expansion
\begin{align}
\dfrac{q_0^4}{\pi} \int_0^\infty \dfrac{ds}{s^2} \dfrac{1}{s-q_0^2} \left ( {\rm Im} \Pi_+(s) + {\rm Im} \Pi_-(s) \right ) + {\it subtractions} & = {\rm Re} \Pi^e_{\rm OPE} (q_0), \\
\dfrac{q_0^5}{\pi} \int_0^\infty \dfrac{ds}{s^{5/2}} \dfrac{1}{s-q_0^2} \left ( {\rm Im} \Pi_+(s) - {\rm Im} \Pi_-(s) \right ) + {\it subtractions} & = {\rm Re} \Pi^o_{\rm OPE} (q_0),
\end{align}
so that sum rules for $D^+$ and $D^-$ yield predictions on the average mass and
the mass splitting; $q_0$ is the time component of the off-shell $D$ meson at rest. It has been suggested that the center of masses $(m_{D^-}+m_{D^+})/2$ decreases slightly in medium~\cite{Hayashigaki:2000es}, accompanied by an enhancement of the splitting
$(m_{D^-}-m_{D^+})$ \cite{Weise:2001wv,pk:Morath2001}. Quantitative uncertainties arise due to the questionable use of a pole + continuum ansatz as well as not well known but important condensates, especially mixed quark-gluon structures.
This point deserves further detailed studies with respect to envisaged investigations within the CBM and PANDA projects at FAIR.

For systems of two heavy quarks $(\bar{Q}Q)$, like $J/\Psi$, one might naively expect a further enhancement of the role of the chiral condensate. However, only the combination $m_Q \condensate{\bar{Q} Q}$ enters the sum rule; in leading order there is no term $\propto \condensate{\bar{q} q}$.
In the heavy quark mass expansion the heavy chiral condensate is related to gluonic condensates \cite{Shifman:1978bx,Generalis:1983hb}
\begin{equation}
m_Q \condensate{\bar{Q} Q} = - \dfrac{1}{12} \condensate{\dfrac{\alpha_s}{\pi} G^2} - \dfrac{\condensate{g_s^3 G^3}}{1440 \pi^2 m_Q^2} 
- \dfrac{1}{30 m_Q^2} \condensate{\dfrac{\alpha_s}{\pi} (D^\mu G^a_{\alpha \mu}) (D_\nu G_a^{\alpha \nu}) }
+ \ldots \, ;
\label{eq:heavyQuarkMassExpansionChiralCondensate}
\end{equation}
the last term in Eq.~(\ref{eq:heavyQuarkMassExpansionChiralCondensate}) is by means of the equations of motion related to four-quark condensates.
A sum rule evaluation for $J/\Psi$ \cite{Klingl:1998sr} shows tiny in-medium effects, conceivable from the lowest dimension gluonic condensate $\condensate{\tfrac{\alpha_s}{\pi} G^2 }$ which is inert against density changes far from the deconfinement boundary.

\section{Summary}
It was pointed out that QCD sum rules relate hadronic properties at finite temperature and/or density
to universal condensates parameterizing the QCD ground state. Among these for $\omega$ meson and nucleon we emphasized
the importance of the density dependence of combined four-quark condensates, which appears to be strong for the $\omega$ case
and weak for the combinations in the nucleon example. In systems containing heavy quarks other condensates contribute dominantly.

Although four-quark condensates can not generally be considered 
as order parameters for chiral symmetry, only specific combinations 
behave like the chiral condensate $\langle \bar{q} q \rangle$ under chiral 
transformations, they nevertheless carry information about the complicated 
structure of the QCD vacuum. This, in summary, signals significant modifications 
of the ground state, and since the condensates represent universal parameters for all hadrons an 
understanding of the $\omega$ and nucleon in-medium properties is inevitable incorporated with the 
density-dependent properties of all low-lying hadrons.

\section*{Acknowledgments}
The work is supported by BMBF and GSI.


\begin{thebibliography}{18}
\expandafter\ifx\csname natexlab\endcsname\relax\def\natexlab#1{#1}\fi
\expandafter\ifx\csname bibnamefont\endcsname\relax
  \def\bibnamefont#1{#1}\fi
\expandafter\ifx\csname bibfnamefont\endcsname\relax
  \def\bibfnamefont#1{#1}\fi
\expandafter\ifx\csname citenamefont\endcsname\relax
  \def\citenamefont#1{#1}\fi
\expandafter\ifx\csname url\endcsname\relax
  \def\url#1{\texttt{#1}}\fi
\expandafter\ifx\csname urlprefix\endcsname\relax\def\urlprefix{URL }\fi
\providecommand{\bibinfo}[2]{#2}
\providecommand{\eprint}[2][]{\url{#2}}

\bibitem[{\citenamefont{Brown and Rho}(1991)}]{Brown:1991kk}
\bibinfo{author}{\bibfnamefont{G.~E.} \bibnamefont{Brown}} \bibnamefont{and}
  \bibinfo{author}{\bibfnamefont{M.}~\bibnamefont{Rho}},
  \bibinfo{journal}{Phys. Rev. Lett.} \textbf{\bibinfo{volume}{66}},
  \bibinfo{pages}{2720} (\bibinfo{year}{1991}).

\bibitem[{\citenamefont{Shifman et~al.}(1979)\citenamefont{Shifman, Vainshtein,
  and Zakharov}}]{Shifman:1978bx}
\bibinfo{author}{\bibfnamefont{M.~A.} \bibnamefont{Shifman}},
  \bibinfo{author}{\bibfnamefont{A.~I.} \bibnamefont{Vainshtein}},
  \bibnamefont{and} \bibinfo{author}{\bibfnamefont{V.~I.}
  \bibnamefont{Zakharov}}, \bibinfo{journal}{Nucl. Phys.}
  \textbf{\bibinfo{volume}{B147}}, \bibinfo{pages}{385} (\bibinfo{year}{1979}).

\bibitem[{\citenamefont{Zschocke et~al.}(2002)\citenamefont{Zschocke,
  K{\"a}mpfer, Pavlenko, and Wolf}}]{Zschocke:2002ic}
\bibinfo{author}{\bibfnamefont{S.}~\bibnamefont{Zschocke}},
  \bibinfo{author}{\bibfnamefont{B.}~\bibnamefont{K{\"a}mpfer}},
  \bibinfo{author}{\bibfnamefont{O.~P.} \bibnamefont{Pavlenko}},
  \bibnamefont{and} \bibinfo{author}{\bibfnamefont{G.}~\bibnamefont{Wolf}}, in
  \emph{\bibinfo{booktitle}{Proceedings of the 40th International Winter
  Meeting on Nuclear Physics}}, edited by
  \bibinfo{editor}{\bibfnamefont{I.}~\bibnamefont{Iori}} \bibnamefont{and}
  \bibinfo{editor}{\bibfnamefont{A.}~\bibnamefont{Moroni}}
  (\bibinfo{address}{Bormio, Italy}, \bibinfo{year}{2002}), pp.
  \bibinfo{pages}{102--111}.

\bibitem[{\citenamefont{Hatsuda and Lee}(1992)}]{Hatsuda:1991ez}
\bibinfo{author}{\bibfnamefont{T.}~\bibnamefont{Hatsuda}} \bibnamefont{and}
  \bibinfo{author}{\bibfnamefont{S.~H.} \bibnamefont{Lee}},
  \bibinfo{journal}{Phys. Rev.} \textbf{\bibinfo{volume}{C46}},
  \bibinfo{pages}{34} (\bibinfo{year}{1992}).

\bibitem[{\citenamefont{Friese}(1999)}]{Friese:1999qm}
\bibinfo{author}{\bibfnamefont{J.}~\bibnamefont{Friese}}
  (\bibinfo{collaboration}{HADES}), \bibinfo{journal}{Prog. Part. Nucl. Phys.}
  \textbf{\bibinfo{volume}{42}}, \bibinfo{pages}{235} (\bibinfo{year}{1999}).

\bibitem[{\citenamefont{Zschocke et~al.}(2003)\citenamefont{Zschocke, Pavlenko,
  and K{\"a}mpfer}}]{Zschocke:2002mp}
\bibinfo{author}{\bibfnamefont{S.}~\bibnamefont{Zschocke}},
  \bibinfo{author}{\bibfnamefont{O.~P.} \bibnamefont{Pavlenko}},
  \bibnamefont{and}
  \bibinfo{author}{\bibfnamefont{B.}~\bibnamefont{K{\"a}mpfer}},
  \bibinfo{journal}{Phys. Lett.} \textbf{\bibinfo{volume}{B562}},
  \bibinfo{pages}{57} (\bibinfo{year}{2003}).

\bibitem[{\citenamefont{Thomas et~al.}(2005{\natexlab{a}})\citenamefont{Thomas,
  Gallmeister, Zschocke, and K{\"a}mpfer}}]{Thomas:2005wc}
\bibinfo{author}{\bibfnamefont{R.}~\bibnamefont{Thomas}},
  \bibinfo{author}{\bibfnamefont{K.}~\bibnamefont{Gallmeister}},
  \bibinfo{author}{\bibfnamefont{S.}~\bibnamefont{Zschocke}}, \bibnamefont{and}
  \bibinfo{author}{\bibfnamefont{B.}~\bibnamefont{K{\"a}mpfer}}
  (\bibinfo{year}{2005}{\natexlab{a}}), \eprint{hep-ph/0501202}.

\bibitem[{\citenamefont{Trnka et~al.}(2005)}]{Trnka:2005ey}
\bibinfo{author}{\bibfnamefont{D.}~\bibnamefont{Trnka}} \bibnamefont{et~al.}
  (\bibinfo{collaboration}{CBELSA/TAPS}), \bibinfo{journal}{Phys. Rev. Lett.}
  \textbf{\bibinfo{volume}{94}}, \bibinfo{pages}{192303}
  (\bibinfo{year}{2005}).

\bibitem[{\citenamefont{Thomas et~al.}(2005{\natexlab{b}})\citenamefont{Thomas,
  Zschocke, and K{\"a}mpfer}}]{Thomas:2005dc}
\bibinfo{author}{\bibfnamefont{R.}~\bibnamefont{Thomas}},
  \bibinfo{author}{\bibfnamefont{S.}~\bibnamefont{Zschocke}}, \bibnamefont{and}
  \bibinfo{author}{\bibfnamefont{B.}~\bibnamefont{K{\"a}mpfer}},
  \bibinfo{journal}{Phys. Rev. Lett.} \textbf{\bibinfo{volume}{95}},
  \bibinfo{pages}{232301} (\bibinfo{year}{2005}{\natexlab{b}}).

\bibitem[{\citenamefont{Zschocke and K{\"a}mpfer}(2004)}]{Zschocke:2004qa}
\bibinfo{author}{\bibfnamefont{S.}~\bibnamefont{Zschocke}} \bibnamefont{and}
  \bibinfo{author}{\bibfnamefont{B.}~\bibnamefont{K{\"a}mpfer}},
  \bibinfo{journal}{Phys. Rev.} \textbf{\bibinfo{volume}{C70}},
  \bibinfo{pages}{035207} (\bibinfo{year}{2004}).

\bibitem[{\citenamefont{Vretenar and Weise}(2004)}]{Vretenar:2003bt}
\bibinfo{author}{\bibfnamefont{D.}~\bibnamefont{Vretenar}} \bibnamefont{and}
  \bibinfo{author}{\bibfnamefont{W.}~\bibnamefont{Weise}},
  \bibinfo{journal}{Lect. Notes Phys.} \textbf{\bibinfo{volume}{641}},
  \bibinfo{pages}{65} (\bibinfo{year}{2004}).

\bibitem[{\citenamefont{Furnstahl et~al.}(1992)\citenamefont{Furnstahl,
  Griegel, and Cohen}}]{Furnstahl:1992pi}
\bibinfo{author}{\bibfnamefont{R.~J.} \bibnamefont{Furnstahl}},
  \bibinfo{author}{\bibfnamefont{D.~K.} \bibnamefont{Griegel}},
  \bibnamefont{and} \bibinfo{author}{\bibfnamefont{T.~D.} \bibnamefont{Cohen}},
  \bibinfo{journal}{Phys. Rev.} \textbf{\bibinfo{volume}{C46}},
  \bibinfo{pages}{1507} (\bibinfo{year}{1992}).

\bibitem[{\citenamefont{Drukarev et~al.}(2003)}]{Drukarev:2003xd}
\bibinfo{author}{\bibfnamefont{E.~G.} \bibnamefont{Drukarev}}
  \bibnamefont{et~al.}, \bibinfo{journal}{Phys. Rev.}
  \textbf{\bibinfo{volume}{D68}}, \bibinfo{pages}{054021}
  (\bibinfo{year}{2003}).

\bibitem[{\citenamefont{Hayashigaki}(2000)}]{Hayashigaki:2000es}
\bibinfo{author}{\bibfnamefont{A.}~\bibnamefont{Hayashigaki}},
  \bibinfo{journal}{Phys. Lett.} \textbf{\bibinfo{volume}{B487}},
  \bibinfo{pages}{96} (\bibinfo{year}{2000}).

\bibitem[{\citenamefont{Weise}(2001)}]{Weise:2001wv}
\bibinfo{author}{\bibfnamefont{W.}~\bibnamefont{Weise}}, in
  \emph{\bibinfo{booktitle}{Structure of Hadrons: 29th International Workshop
  on Gross Properties of Nuclei and Nuclear Excitations}}
  (\bibinfo{address}{Hirschegg, Austria}, \bibinfo{year}{2001}),
  \urlprefix\url{http://theory.gsi.de/hirschegg/proceedings.html}.

\bibitem[{\citenamefont{Morath}(2001)}]{pk:Morath2001}
\bibinfo{author}{\bibfnamefont{P.}~\bibnamefont{Morath}},
  \bibinfo{type}{{Ph.D.} dissertation}, \bibinfo{school}{TU M{\"u}nchen}
  (\bibinfo{year}{2001}).

\bibitem[{\citenamefont{Generalis and Broadhurst}(1984)}]{Generalis:1983hb}
\bibinfo{author}{\bibfnamefont{S.~C.} \bibnamefont{Generalis}}
  \bibnamefont{and} \bibinfo{author}{\bibfnamefont{D.~J.}
  \bibnamefont{Broadhurst}}, \bibinfo{journal}{Phys. Lett.}
  \textbf{\bibinfo{volume}{B139}}, \bibinfo{pages}{85} (\bibinfo{year}{1984}).

\bibitem[{\citenamefont{Klingl et~al.}(1999)\citenamefont{Klingl, Kim, Lee,
  Morath, and Weise}}]{Klingl:1998sr}
\bibinfo{author}{\bibfnamefont{F.}~\bibnamefont{Klingl}},
  \bibinfo{author}{\bibfnamefont{S.-s.} \bibnamefont{Kim}},
  \bibinfo{author}{\bibfnamefont{S.~H.} \bibnamefont{Lee}},
  \bibinfo{author}{\bibfnamefont{P.}~\bibnamefont{Morath}}, \bibnamefont{and}
  \bibinfo{author}{\bibfnamefont{W.}~\bibnamefont{Weise}},
  \bibinfo{journal}{Phys. Rev. Lett.} \textbf{\bibinfo{volume}{82}},
  \bibinfo{pages}{3396} (\bibinfo{year}{1999}), \bibinfo{note}{erratum: Phys.
  Rev. Lett. {\bf 83}, 4224 (1999)}.

\end{thebibliography}

\end{document}